\documentclass[pre,aps,floats,twocolumn,superscriptaddress,floatfix]{revtex4}
\usepackage{graphicx,amssymb,amsmath,ifthen,rotating}
\usepackage[active]{srcltx}
\begin{document}
\title{Statistically interacting vacancy particles}  
\author{Benaoumeur Bakhti}
\affiliation{
  Fachbereich Physik,
  Universit\"at Osnabr\"uck,
  D-49076 Osnabr\"uck, Germany}
      \author{Michael Karbach}
\affiliation{
Fachgruppe Physik,
  Bergische Universit{\"{a}}t Wuppertal,
  D-42097 Wuppertal, Germany}
  \author{Philipp Maass}
\affiliation{
  Fachbereich Physik,
  Universit\"at Osnabr\"uck,
  D-49076 Osnabr\"uck, Germany}
  \author{Mohammad Mokim}
\affiliation{
  Department of Physics,
  University of Rhode Island,
  Kingston RI 02881, USA}
\author{Gerhard M{\"{u}}ller}
\affiliation{
  Department of Physics,
  University of Rhode Island,
  Kingston RI 02881, USA}

\begin{abstract}
  The equilibrium statistical mechanics of one-dimensional lattice gases with interactions of arbitrary range and shape between first-neighbor atoms is solved exactly on the basis of statistically interacting vacancy particles.
  Two sets of vacancy particles are considered.
  In one set all vacancies are of one-cell size.
  In the other set the sizes of vacancy particles match the separation between atoms.
  Explicit expressions are obtained for the Gibbs free energy and the distribution of spaces between atoms at thermal equilibrium.
  Applications to various types of interaction potentials are discussed, including long-range potentials that give rise to phase transitions. 
  Extensions to hard rod systems are straightforward and are shown to agree with existing results for lattice models and their continuum limits.
\end{abstract}

\maketitle

%
\section{Introduction}\label{sec:intro}
%
Lattice-gas models are widely used in statistical mechanics with a host of applications across many fields of research within physics and beyond \cite{Reic98}.
Exact solutions are ubiquitous in $\mathcal{D}=1$ \cite{LM66}, rare and precious in $\mathcal{D}=2$ \cite{Onsa44, Yang52, Baxt82}, and seemingly out of reach in $\mathcal{D}=3$.

The transfer-matrix analysis is tailor-made for lattice gases in $\mathcal{D}=1$ with homogeneous short-range interactions \cite{KW41}. 
However, the analytic or computational costs increase exponentially with the range of the interatomic coupling.
Takahashi \cite{Taka42} found a way of avoiding that runaway price for interactions of arbitrary range between successive atoms on a line. 

The statistical mechanical analysis presented in this work for lattice systems with first-neighbor interactions of arbitrary range and shape is based on a different approach. 
The exact solutions emerging from it are shown to be equivalent to a lattice version of Takahashi's work and to have potential for significant extensions in scope.

Our method is constructed within the conceptual framework of fractional statistics, originally developed in the context of quantum many-body systems \cite{Hald91a, Wu94, Isak94, Anghel} and later adapted to the classical mechanics of statistically interacting particles with shapes \cite{LVP+08, copic, picnnn, pichs, GKLM13}. 
The basic idea, in a nutshell, is to replace fundamental degrees of freedom with simple statistics and couplings of any range and strength by equivalent degrees of freedom with complex statistics but no couplings.
The thermodynamic Bethe ansatz \cite{Taka99} has been shown \cite{BW94} to be one realization  of this general idea \cite{PMK07}.
Here we present a different realization that is suitable for classical systems.

\begin{figure}[b]
  \begin{center}
 \includegraphics[width=79mm]{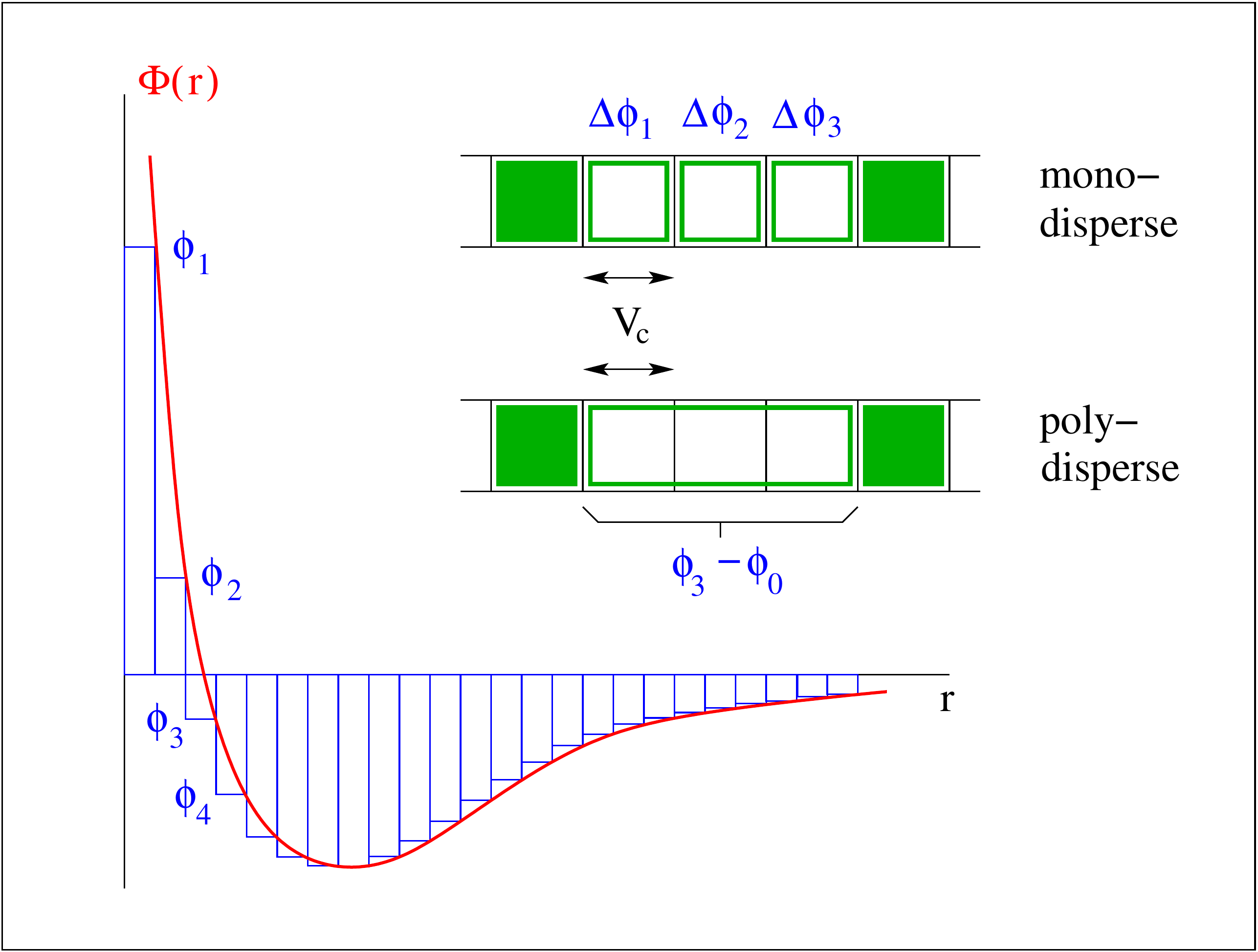}
\end{center}
\caption{(Color online) Discretization of a potential $\Phi(r)$ into a sequence of values $\phi_m$ at distances $r=mV_c$, with cell size $V_c$.
Monodisperse particles (of unit size) have excitation energies $\epsilon_m=pV_c+\Delta\phi_m$, where $\Delta\phi_m=\phi_m-\phi_{m-1}$ and $p$ is the ambient pressure.
Polydisperse particles of size $m$ have excitation energies $\epsilon_m=mpV_c+\phi_m-\phi_0$.}
  \label{fig:fig1}
\end{figure}

In the lattice gas under scrutiny here, the fundamental degrees of freedom are cells that can be vacant or singly occupied. 
The system is understood to be exposed to ambient pressure $p$ and to extend from the first to the last of $N_A$ occupied cells.
The coupling between first-neighbor occupied cells separated by any distance is described by a pair potential $\Phi(r)$ discretized as illustrated in Fig.~\ref{fig:fig1}.
The work required to separate two atoms in adjacent cells against pressure and interaction force is interpreted as the excitation energy of \emph{vacancy particles} associated with the vacant cells thus created between the two atoms.

In Sec.~\ref{sec:comb} we work out the combinatorial exclusion statistics of two sets of vacancy particles.
In one set all particles have the size of one vacant cell.
They coexist in nested configurations.
The particles of the other set have sizes equal to the clusters of vacant cells between successive occupied cells.

In Sec.~\ref{sec:stat-mech} we present the exact statistical mechanical analysis for a generic interaction potential based on both sets of vacancy particles.
It includes expressions for the Gibbs free energy and the distribution of spaces between occupied cells, which is a measure for correlations between occupied cells.
In Sec.~\ref{sec:app} we work out applications to potentials of various shapes and ranges.
The versatility of the approach is demonstrated in the process.
Its promise for significant extensions is briefly discussed in Sec.~\ref{sec:ext}.

%
\section{Combinatorics and energetics}\label{sec:comb}
%
Statistically interacting particles are free of interaction energies but their placement into the system is governed by rules derived from a generalized Pauli principle.
How is the number $d_m$ of states accessible to a particle of species $m$ affected if the number $N_{m'}$ of particles from any species $m'$ is changed?
The answer proposed by Haldane in a particular quantum many-body context \cite{Hald91a},
\begin{equation}\label{eq:1} 
\Delta d_m=-\sum_{m'}g_{mm'}\Delta N_{m'},
\end{equation}
which introduces an array of rational coefficients $g_{mm'}$, has turned out to be widely adaptable.

In the lattice-gas context, we are dealing with $M$ species of vacancy particles that are thermally excited from a unique pseudo-vacuum with energy $E_\mathrm{pv}$ (state with all $N_A$ atoms in one cluster). 
The energy and multiplicity of many-body states with given particle content $\{N_m\}$ can then be expressed in the form
\begin{equation}\label{eq:2} 
E(\{N_m\})=E_{\mathrm{pv}}+\sum_{m=1}^M N_m\epsilon_m,
\end{equation}
\begin{subequations}\label{eq:3} 
\begin{align}\label{eq:3a} 
W(\{N_m\}) &=\prod_{m=1}^M\left(\begin{array}{c}
d_m+N_m-1 \\ N_m\end{array}\right), \\ \label{eq:3b} 
 d_m &=A_m-\sum_{m'=1}^M g_{mm'}(N_{m'}-\delta_{mm'}).
\end{align}
\end{subequations}
The $\epsilon_m$ are particle excitation energies, the $A_m$ are capacity constants, and the $g_{mm'}$ are statistical interaction coefficients.
Our methodology uses an ensemble with a fixed number $N_A$ of occupied cells.
The number of vacant cells at given pressure is fluctuating.

In the following we introduce two sets of vacancy particles from which we derive equivalent results for the statistical mechanical properties of lattice gases with an arbitrary interaction potential $\Phi(r)$.
Depending on the shape and range of $\Phi(r)$, the use of one set may offer advantages over the use of the other.

\subsection{Monodisperse vacancy particles}\label{sec:mono}
In this set all vacancy particles have size $V_c$ and are placed in nested configurations between occupied cells. 
Each vacancy particle placed increases the distance between two first-neighbor occupied cells by $V_c$.
If the range of the interatomic coupling is $(M-1)V_c$ the work performed in the process is 
\begin{equation}\label{eq:4}
\epsilon_m=\left\{\begin{array}{ll} pV_c+\Delta\phi_m, & m=1,\ldots,M-1 \\ pV_c, & m=M 
\end{array} \right.,
\end{equation}
where $\Delta\phi_m=\phi_m-\phi_{m-1}$ (see Fig.~\ref{fig:fig1}).
We have $M$ species from three categories in the taxonomy of Ref.~\cite{copic}: one species $(m=1)$ of \emph{hosts}, $M-2$ species of \emph{hybrids}, and one species $(m=M)$ of \emph{tags}.
Hosts cannot be hosted, tags cannot host any particles from a different species, hybrids can do both.

The first particle to be placed between adjacent occupied cells is a host, the following ones are hybrids from species $m=2,3,\ldots$ in exactly that sequence. 
Any particle added in excess of $M-1$ vacant cells between two first-neighbor occupied cells is a tag.

The pseudo-vacuum has capacity for hosts, not for hybrids or tags. 
We have
\begin{equation}\label{eq:5}
A_m=\left\{\begin{array}{ll} N_A-1, & m=1 \\ 0, & m=2,\ldots,M 
\end{array} \right..
\end{equation}
Capacity for hybrids of species $m=2$ must be generated by first placing hosts.
Placing hybrids of species $m=2$, in turn, creates capacity for placing hybrids of species $m=3$ etc.

The diagonal interaction coefficients,
\begin{equation}\label{eq:6}
g_{mm}=\left\{\begin{array}{ll} 1, & m=1,\ldots,M-1\\ 0, & m=M 
\end{array} \right.,
\end{equation}
reflect the limited capacity of hosts and hybrids (one from each species between any two occupied cells) and the unlimited capacity for tags (if hybrids from species $m=M-1$ are present).
The only nonzero off-diagonal interaction coefficients are negative,
\begin{equation}\label{eq:7} 
g_{m,m-1}=-1,\quad m=2,\ldots,M,
\end{equation}
and reflect the dynamic generation of capacity for hybrids and tags.

The number of monodisperse particles that can be placed into the system is unlimited.
The number of species is finite and depends on the range of $\Phi(r)$.

\subsection{Polydisperse vacancy particles}\label{sec:poly}
Vacancy particles from this set come in sizes $mV_c$, $m=1,2,\ldots$
No more than one particle can be placed between occupied cells.
The number of particles cannot exceed $N_A-1$. 
But the variety of species is infinite.
The limit $M\to\infty$ is implied in Eqs.~(\ref{eq:2}) and (\ref{eq:3}).

Unable to host or being hosted, these particles belong to the category of \emph{compacts} \cite{copic}.
Placing a vacancy particle of species $m$ means increasing the distance between two occupied cells from zero to $mV_c$.
The work performed in the process is (see Fig.~\ref{fig:fig1})
\begin{equation}\label{eq:8} 
\epsilon_m=mpV_c+\phi_m-\phi_0.
\end{equation}
The pseudo-vacuum has equal capacity for compacts of all sizes:
\begin{equation}\label{eq:9} 
A_m=N_A-1,\quad m=1,2,\ldots
\end{equation}
Placing a compact of species $m$ reduces the number of open slots for placing further compacts of any species $m'$.
This particular statistical interaction is encoded in the interaction coefficients as follows:
\begin{equation}\label{eq:10} 
g_{mm'}=\left\{\begin{array}{ll} 1, & m'\geq m\\ 0, & m'<m
\end{array} \right..
\end{equation}
The expected symmetry of the $g_{mm'}$ is concealed in the attribute that exchanging $m$ and $m'$ on the right of Eq.~(\ref{eq:10}) produces the same multiplicity $W(\{N_m\})$.

%
\section{Statistical Mechanics}\label{sec:stat-mech}
%
The partition function,
\begin{equation}\label{eq:11} 
Z=\sum_{\{N_m\}}W(\{N_m\})\exp\big(-\beta E(\{N_m\})\big),
\end{equation}
depends on energy (\ref{eq:2}) and multiplicity (\ref{eq:3}) with ingredients $\epsilon_m, A_m, g_{mm'}$.
The interaction potential $\Phi(r)$ is encoded in the $\phi_m$, which are contained in the $\epsilon_m$.

The extremum principle from which the thermal equilibrium macrostate in the thermodynamic limit follows was carried out for generic systems \cite{Wu94, Isak94}.
It yields the partition function in the form \cite{LVP+08, copic, picnnn, pichs},
\begin{equation}\label{eq:12} 
Z=\prod_{m=1}^M\big(1+w_m^{-1}\big)^{A_m},
\end{equation}
where the (real, positive) $w_m$ are solutions of the coupled nonlinear equations,
\begin{equation}\label{eq:13} 
e^{\beta\epsilon_m}=(1+w_m)\prod_{m'=1}^M \big(1+w_{m'}^{-1}\big)^{-g_{m'm}}.
\end{equation}
The numbers of vacancy particles (per occupied cell) from species $m$, $\bar{N}_m\doteq\langle N_m\rangle/N_A$, are derived from the coupled linear equations \cite{Wu94,Isak94},
\begin{equation}\label{eq:14} 
w_m\bar{N}_m+\sum_{m'=1}^Mg_{mm'}\bar{N}_{m'} =\bar{A}_m,
\end{equation}
where $\bar{A}_m\doteq A_m/N_A$.
The $\bar{N}_m$ can be interpreted as population densities of particles from species $m$.
For compacts, hosts, and hybrids they are restricted to $0\leq\bar{N}_m\leq1$.
Tag population densities have no upper limit.
The derivation of Eqs.~(\ref{eq:12})-(\ref{eq:14}) from Eq.~(\ref{eq:11}) is outlined in Appendix~\ref{sec:parti}.

In our context, the thermodynamic potential inferred from $Z$ is the Gibbs free energy.
Its natural independent variables $T,p,N_A$ are contained in the $w_m,\epsilon_m,A_m$, respectively.
From the scaled Gibbs free energy,
\begin{equation}\label{eq:15} 
 \bar{G}(T,p)=-\lim_{N_A\to\infty}N_A^{-1}k_BT\ln Z=\bar{U}-T\bar{S}+p\bar{V},
\end{equation}
we can derive expressions for the scaled entropy $\bar{S}$, the scaled excess volume, 
\begin{equation}\label{eq:64} 
\bar{V}\doteq\frac{V-V_\mathrm{ex}}{N_A}=\frac{V}{N_A}-V_c,
\end{equation}
with exclusion volume $V_\mathrm{ex}\doteq N_AV_c$, the scaled internal energy $\bar{U}=U/N_A$, and various response functions.
The excess volume can also be calculated from the population densities of vacancy particles via
\begin{equation}\label{eq:16} 
 \bar{V}=\sum_{m=1}^M\bar{N}_mv_m,
\end{equation}
where $v_m=V_c$ ($v_m=mV_c$) for the set of monodisperse (polydisperse) vacancy particles.
Likewise, the entropy can be derived via $S=k_B\ln W$, yielding \cite{Isak94,picnnn}
\begin{subequations}\label{eq:17}
\begin{align}\label{eq:17a}
\bar{S}(\{\bar{N}_m\}) &= k_B\sum_{m=1}^M
\Big[\big(\bar{N}_{m}+\bar{Y}_m\big)\ln\big(\bar{N}_m+\bar{Y}_m\big) \nonumber \\
&-\bar{N}_m\ln \bar{N}_m -\bar{Y}_m\ln \bar{Y}_m\Big], \\ \label{eq:17b}
 \bar{Y}_m &\doteq \bar{A}_m-\sum_{m'=1}^Mg_{mm'} \bar{N}_{m'}.
\end{align}
\end{subequations}

\subsection{Hosts, hybrids, and tags}\label{sec:ho-hy-ta}
Here we present an exact solution in general form using the monodisperse vacancy particles with specifications (\ref{eq:4})-(\ref{eq:7}). 
Equations (\ref{eq:13}) can be solved sequentially,
\begin{align}\label{eq:18} 
& w_M=e^{\beta\epsilon_M}-1;\nonumber \\
& w_m=e^{\beta\epsilon_m}\frac{w_{m+1}}{1+w_{m+1}},\quad m=M-1,\ldots,1,
\end{align}
yielding the free energy,
\begin{equation}\label{eq:19} 
\beta\bar{G}(T,p)=-\ln(1+w_1^{-1}).
\end{equation}
The population densities of host, hybrids, and tag are then derived from the sequential solution of Eqs.~(\ref{eq:14}):
\begin{subequations}\label{eq:20} 
\begin{align}
&\bar{N}_m=\prod_{m'=1}^m(1+w_{m'})^{-1},\quad m=1,\ldots,M-1; \\
&\bar{N}_M=w_M^{-1}\prod_{m'=1}^{M-1}(1+w_{m'})^{-1}.
\end{align}
\end{subequations}

\subsection{Compacts}\label{sec:comp}
An equivalent but structurally different solution uses the polydisperse vacancy particles with specifications (\ref{eq:8})-(\ref{eq:10}).
Equations (\ref{eq:13}), now infinite in number, can be solved in closed form:
\begin{equation}\label{eq:21} 
w_m=e^{\beta\epsilon_m}\left[1+\sum_{m'=1}^{m-1}e^{-\beta\epsilon_{m'}}\right],\quad m=1,2,\ldots
\end{equation}
The free energy, inferred via (\ref{eq:12}) and (\ref{eq:15}), becomes
\begin{equation}\label{eq:22} 
\beta\bar{G}(T,p)=-\ln\left(1+\sum_{m=1}^\infty e^{-\beta\epsilon_m}\right).
\end{equation}

It turns out that all thermodynamic quantities of interest can be expressed in terms of the functions
\begin{equation}\label{eq:30} 
 B_{lk}(T,p)\doteq\sum_{m=0}^\infty m^l(\beta\epsilon_m)^ke^{-\beta\epsilon_m},
\end{equation}
with $\epsilon_0=0$ implied.
We thus obtain the following explicit or parametric expressions for the excess volume $\bar{V}$, the density of occupied cells $\rho\doteq N_AV_c/V$, entropy $\bar{S}$, internal energy $\bar{U}$, heat capacity $\bar{C}_p\doteq T(\partial\bar{S}/\partial T)_p$, compressibility $\kappa_T\doteq-\bar{V}^{-1}(\partial\bar{V}/\partial p)_T$, and thermal expansivity $\alpha_p\doteq\bar{V}^{-1}(\partial\bar{V}/\partial T)_p$:
\begin{subequations}\label{eq:37} 
\begin{equation}\label{eq:37a} 
 \frac{\bar{V}}{V_c} = \frac{1}{\rho}-1=\frac{B_{10}}{B_{00}},\quad
 \frac{\bar{S}}{k_B}=\ln B_{00}+\frac{B_{01}}{B_{00}},
\end{equation}
\begin{equation}\label{eq:37b} 
\bar{U}=k_BT\frac{B_{01}}{B_{00}}-pV_c\frac{B_{10}}{B_{00}},\quad
 \frac{\bar{C}_p}{k_B}= \frac{B_{02}}{B_{00}}-\left[ \frac{B_{01}}{B_{00}}\right]^2,
\end{equation}
\begin{equation}\label{eq:37c} 
 \frac{\kappa_T}{\beta V_c}=\frac{B_{20}}{B_{10}}-\frac{B_{10}}{B_{00}},\quad
 T\alpha_p=\frac{B_{11}}{B_{10}}-\frac{B_{01}}{B_{00}}.
\end{equation}
\end{subequations}

The population densities of compacts, calculated from Eqs.~(\ref{eq:14}), acquire the form
\begin{equation}\label{eq:23} 
\bar{N}_m=\frac{e^{-\beta\epsilon_m}}{B_{00}},\quad m=0,1,2,\ldots,
\end{equation}
where $\bar{N}_0$ (with $\epsilon_0=0$ implied) is the probability that two occupied cells are in contact.
In Appendix~\ref{sec:appb} we show how the $\bar{N}_m$ are related to spatial correlations of occupied cells.

The approach based on monodisperse vacancy particles is well suited for short-range interaction potentials. 
Only a few species are then needed.
The use of polydisperse vacancy particles is more natural for long-range interactions.
The number of species is always infinite.
In some instances this offers advantages even for short-range interactions.

%
\section{Applications}\label{sec:app}
%
In the following we consider various interaction potentials to showcase the versatility of the method. 
For a concise notation we set
\begin{equation}\label{eq:24} 
 x_p=e^{K_p}\doteq e^{\beta pV_c},\quad x_u=e^{K_u}\doteq e^{\beta u},
\end{equation}
where $u$ is a unit of potential energy. 
Positive (negative) $u$ mean attractive (repulsive) interaction forces throughout this work.

\subsection{Contact interaction}\label{sec:cont}
In the ideal lattice gas (ILG), where the hardcore exclusion is the only interaction between occupied cells, the free energy becomes $\beta\bar{G}(T,p)=\ln(1-x_p^{-1})$, the monodisperse vacancy particles are bosonic in nature, $\bar{N}_1=(x_p-1)^{-1}$, and the polydisperse vacancy particles are Pascal distributed, $\bar{N}_m=x_p^{-m}(1-x_p^{-1})$.

\begin{figure}[b]
  \begin{center}
 \includegraphics[width=42mm]{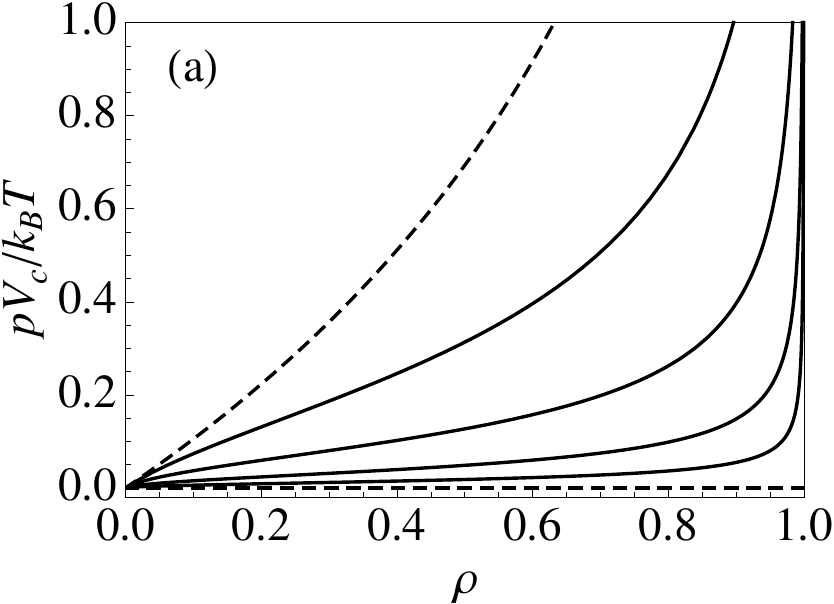}\hspace{2mm}%
 \includegraphics[width=42mm]{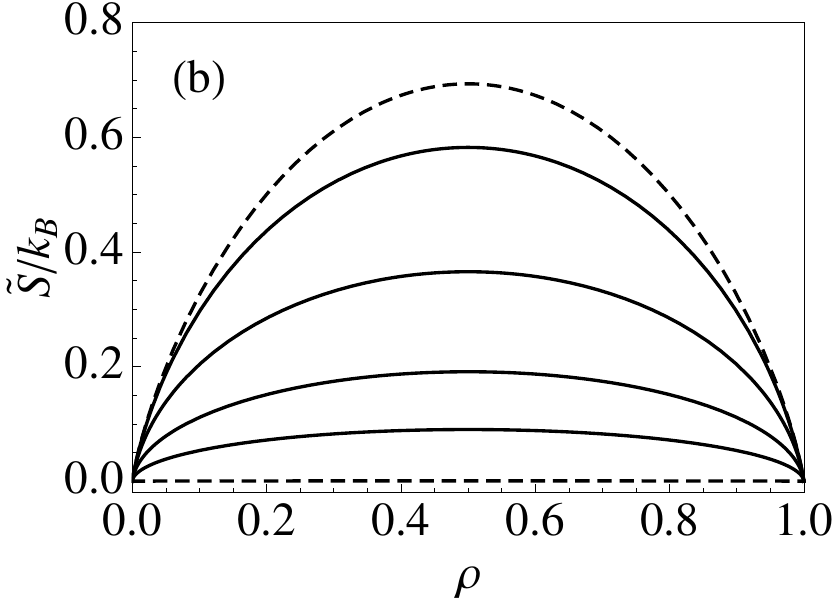}
  \includegraphics[width=42mm]{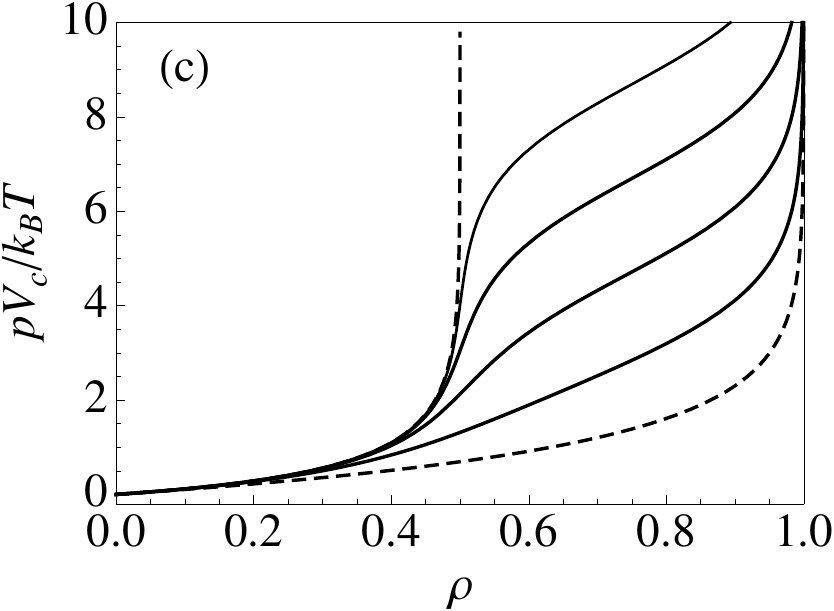}\hspace{2mm}%
  \includegraphics[width=42mm]{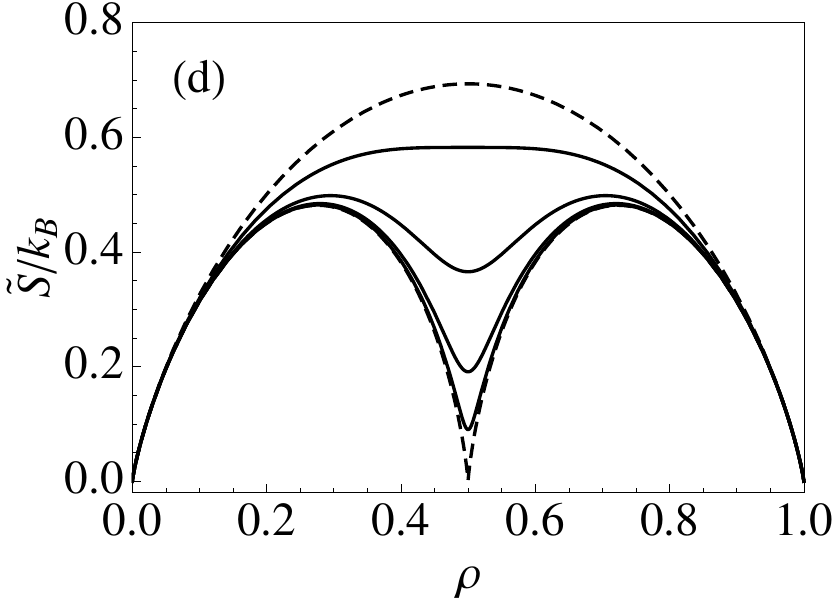}
\end{center}
\caption{Equation of state (left) and entropy (right) versus density for various (scaled) values of attractive (top) and repulsive (bottom) contact potentials 
$\beta u=0, \pm2, \pm4, \pm6, \pm8,\pm\infty $.
The dashed lines represent zero or infinite coupling.}
  \label{fig:fig2}
\end{figure}

The presence of a contact potential ($\Delta\phi_1=u$, $\Delta\phi_m=0$ for $m\geq2$, Ising lattice gas) yields the free energy,
\begin{equation}\label{eq:27} 
\beta\bar{G}(T,p)=-\ln\left(1+\frac{1}{x_u(x_p-1)}\right).
\end{equation}
The equation of state and the entropy inferred from (\ref{eq:27}),
\begin{equation}\label{eq:47} 
 \frac{\bar{V}}{V_c}=\frac{1}{\rho}-1=\frac{x_p}{(x_p-1)[x_u(x_p-1)+1]},
\end{equation}
\begin{equation}\label{eq:31} 
 \frac{\bar{S}}{k_B}=\ln\left(1+\frac{1}{x_u(x_p-1)}\right)
 +\frac{x_p(K_p+K_u)-K_u}{(x_p-1)[x_u(x_p-1)+1]},
\end{equation}
are shown in Fig.~\ref{fig:fig2}.
In panels (a) and (c) we plot $pV_c/k_BT$ versus density $\rho$, and in panels (b) and (d) the entropy per cell, $\tilde{S}=\rho\bar{S}$, versus $\rho$.

In the ILG limit $(u=0)$, we have $pV_c/k_BT=-{\ln(1-\rho)}$. 
The linear behavior at $\rho\ll1$ represents a classical ideal gas. 
The logarithmic divergence for $\rho\to1$ reflects the effect of the hardcore repulsion, manifest in the single-occupancy requirement. 
The characteristic structure of the entropy, $\tilde{S}/k_B=-\rho\ln\rho-(1-\rho)\ln(1-\rho)$, is another signature of hardcore repulsion.

Attractive coupling $u>0$ enhances atomic clustering, which lowers the pressure overall and reduces the entropy overall. 
Infinitely strong attraction leads to zero pressure and zero entropy at all $\rho<1$.

The impact of a repulsive contact interaction depends more strongly on atomic crowding.
At $\rho\ll1$ the occupied cells can avoid each other easily.
Short-range repulsion, unlike short-range attraction, has almost no effect on pressure and entropy.
As the occupied cells approach half filling at sufficiently strong repulsion, the pressure rises fast and the entropy drops to a local minimum.
Infinitely strong repulsion makes the pressure, $\beta pV_c =\ln(1-\rho) -\ln(1-2\rho)$, diverge for $\rho\to\frac{1}{2}$.
The macrostate at $\rho=\frac{1}{2}$ has zero entropy. 
Vacant and occupied cells are frozen into an alternating sequence.

The density of vacuum elements (probability of adjacent occupied cells),
\begin{equation}\label{eq:65} 
 \bar{N}_0=\frac{x_u(x_p-1)}{x_u(x_p-1)+1},
\end{equation}
approaches unity for $u\to+\infty$ (strong attraction) and zero for $u\to-\infty$ (strong repulsion).
The distribution of spacings between occupied cells,
\begin{equation}\label{eq:28} 
 \bar{N}_m=\frac{x_p^{-m}(x_p-1)}{x_u(x_p-1)+1},\quad m=1,2,\ldots,
\end{equation}
tails off exponentially with size and is not affected by the interaction except for the overall normalization factor.
This last attribute is not shared by the pair correlations of occupied cells (see Appendix~\ref{sec:appb}).

\subsection{Square-well potential}\label{sec:rec-wel}
We next consider the interaction potential, $\Phi(r)=u\theta(r-r_0)$, a well of depth $u$ and range $r_0=(M-1)V_c$.
For monodisperse vacancy particles with excitation energies, $\epsilon_m=pV_c+u\delta_{m, M-1}$, the solutions of Eqs.~(\ref{eq:18}) become $w_m=x_p-1$ and
\begin{equation}\label{eq:32} 
 w_{M-l}=\frac{x_p-1}{1-(1-x_u^{-1})x_p^{1-l}},\quad l=1,\ldots,M-1.
\end{equation}
For the free energy (\ref{eq:19}) we infer
\begin{equation}\label{eq:39} 
 \beta\bar{G}(T,p)=-\ln\left(\frac{1}{1-x_p^{-1}}-\frac{(x_u-1)x_p^{2-M}}{x_u(x_p-1)}\right).
\end{equation}
If instead we evaluate the sum in (\ref{eq:22}) with excitation energies, $\epsilon_m=mpV_c$ for $m=1,\ldots,M-1$ and $\epsilon_m=mpV_c+u$ for $m=M-1,M,\ldots$, for polydisperse particles we again arrive at (\ref{eq:39}).
The ILG is recovered from (\ref{eq:39}) for $M=1, u=0$ and the case of a contact potential for $M=2, u\neq0$.

\begin{figure}[t]
  \begin{center}
 \includegraphics[width=41.5mm]{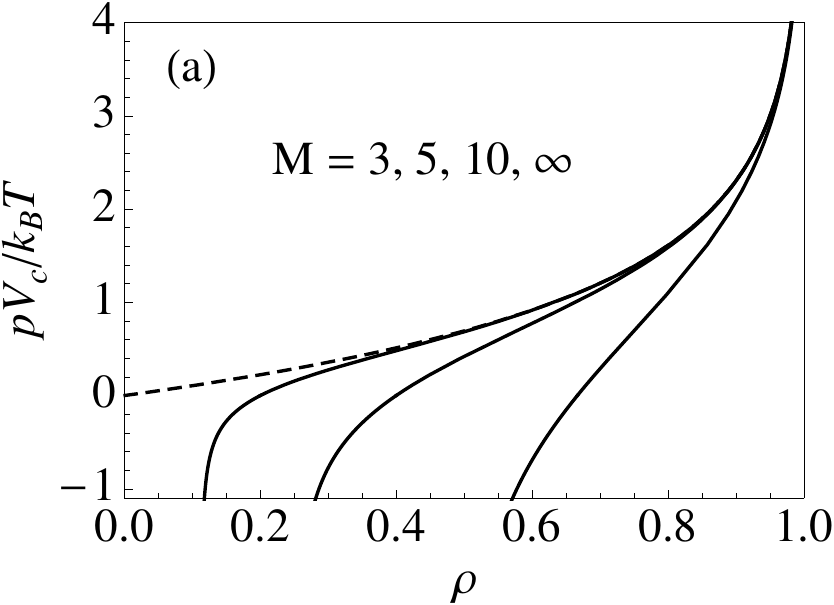}\hspace{2mm}%
 \includegraphics[width=42.5mm]{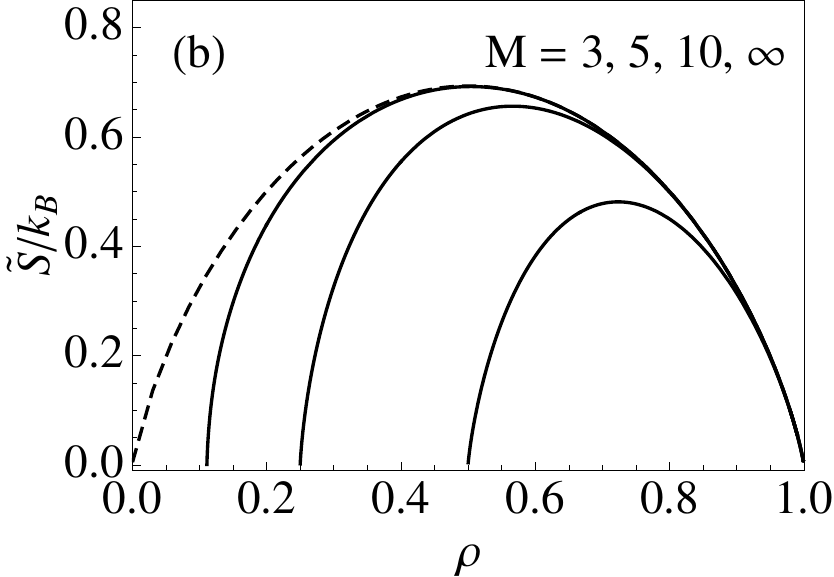}
\end{center}
\caption{ (a) Equation of state and (b) entropy versus density for square-well potentials of various widths $(M-1)V_c$ in the limit of infinite depth. The ILG results (dashed lines) are recovered in the limit $M\to\infty$.}
  \label{fig:fig5}
\end{figure}

The limit $u\to\infty$ of an infinitely deep well (with ${M>2}$) represents an array of tethered beads.
The beads are randomly placed with the constraint that the distance between first neighbors must not exceed $r_0$.
The equation of state and the entropy for this system are
\begin{equation}\label{eq:58} 
 \frac{pV_c}{k_BT}=\ln x_p,\quad 
 \rho=\frac{(x_p-1)(x_p^{M-1}-1)}{x_p^M-M(x_p-1)-1},
\end{equation}
\begin{equation}\label{eq:59} 
 \frac{\tilde{S}}{k_B}=\rho K_p\left[\frac{1}{x_p-1}-\frac{M-1}{x_p^{M-1}-1}\right]
 +\rho\ln\left(\frac{1-x_p^{1-M}}{1-x_p^{-1}}\right),
\end{equation}
for $M=3,4,\ldots$ in a parametric representation.
Plots of $pV_c/k_BT$ and $\tilde{S}/k_B$ versus $\rho$ are shown in Fig.~\ref{fig:fig5}.

The maximum-distance constraint restricts densities to $\rho_\mathrm{min}\leq\rho\leq1$ with $\rho_\mathrm{min}=1/(M-1)$.
The state at minimum density has zero entropy.
For $\rho$ well above $\rho_\mathrm{min}$ the maximum-distance constraint becomes irrelevant and the curves reflect ILG behavior.

When we lower $\rho$ from the ILG regime toward $\rho_\mathrm{min}$ we first observe the entropy to reach a maximum and then the pressure to go negative.
Zero pressure is realized for $\rho_0=2/M$.
Equilibrium states at lower densities only exist under tension.

When we allow the volume to contract an amount $mN_AV_c$ from its maximum value at $\rho_\mathrm{min}$, we allow an average size $mV_c$ of position fluctuation to atoms.
At $\rho_\mathrm{min}$ the distance between atoms is $\sim MV_c$.
Hence the hardcore repulsion is expected to become relevant, i.e to offset the tension, when $m\sim M/2$, which explains the value of $\rho_0$.

\subsection{Uniform attractive force}\label{sec:tri-wel}
The potential $\Phi(r)=-u_0+ur/V_c$ with $u_0={(M-1)u}$ represents a uniform, attractive force $u/V_c$ of range $(M-1)V_c$ between first-neighbor atoms.
Equations~(\ref{eq:18}) for host $(m=1)$ and hybrids $(m=2,\ldots,M-1)$ with excitation energies $\epsilon_m=pV_c+u$, and tag $(m=M)$ with $\epsilon_M=pV_c$ lead to the solution $w_M=x_p-1$ and
\begin{equation}\label{eq:25} 
w_{M-l} =\frac{(x_px_u)^{l-1}}{\displaystyle \frac{(x_px_u)^{l-1}-1}{x_px_u-1}+w_{M-1}^{-1}},
\quad l=1,\ldots,M-1.
\end{equation}
The free energy (\ref{eq:19}) becomes
\begin{equation}\label{eq:26} 
\beta\bar{G}(T,p)=-\ln\left(\frac{1+(x_px_u)^{1-M}\left[\displaystyle\frac{1-x_u^{-1}}{x_{p}-1}\right]}
{1-(x_px_u)^{-1}}\right).
\end{equation}
This solution is also obtained from Eqs.~(\ref{eq:21}) and (\ref{eq:22}) for compacts with excitation energies $\epsilon_m=mpV_c+mu$, $m<M$ and $\epsilon_m=mpV_c+(M-1)u$, $m\geq M$.

Uniform attractive forces of infinite range (equivalent-neighbor models) produce realizations of mean-field theory.
Fluctuations are strongly suppressed and phase transitions are independent of space dimensionality \cite{Stan87}.

The limit $M\to\infty$ of our model with a uniform attractive force is not an equivalent-neighbor model. 
The interaction force between occupied cells is independent of distance but it acts only between first neighbors. 
Thermal fluctuations are stronger by comparison.
Expression (\ref{eq:26}) for the Gibbs free energy simplifies into
\begin{equation}\label{eq:33} 
 \beta\bar{G}(T,p)=\ln\Big(1-(x_px_u)^{-1}\Big)
\end{equation}
and the equation of state becomes
\begin{equation}\label{eq:34} 
 \frac{pV_c}{u}=\frac{k_BT}{u}\ln\left(1+\frac{V_c}{\bar{V}}\right)-1.
\end{equation}
If the system is confined to a box of (scaled) volume $\bar{V}_b$, it undergoes a transition at temperature
\begin{equation}\label{eq:48} 
 k_BT^*=\frac{u}{\ln\big(1+V_c/\bar{V}_b\big)}.
\end{equation}

In the phase at $T>T^*$ we have $p>0$ and $\bar{V}=\bar{V}_b$.
Lowering $T$ toward $T^*$ reduces $p$ linearly at constant $\bar{V}$.
At $T^*$ the pressure vanishes.
A further reduction of $T$ keeps the pressure at zero and decreases the volume below that of the box:
\begin{equation}\label{eq:66} 
 \frac{\bar{V}}{V_c}=\frac{1}{x_u-1}.
\end{equation}
The occupied cells are now self confined. 
The configurational entropy remains nonzero at $T<T^*$.
The entropy expression derived from (\ref{eq:33}),
\begin{equation}\label{eq:35} 
 \frac{\bar{S}}{k_B}=\left(1+\frac{\bar{V}}{V_c}\right)\ln\left(1+\frac{\bar{V}}{V_c}\right)
 -\frac{\bar{V}}{V_c}\ln\frac{\bar{V}}{V_c},
\end{equation}
is a function of $\bar{V}$ alone, implying that it is constant for $T>T^*$ and decreases monotonically as $T$ drops below $T^*$, reaching zero at $T=0$.
The distribution of spacings between occupied cells remains geometric at all $T$, $\bar{N}_m=\rho(1-\rho)^m$, with $\rho=1-x_u^{-1}$ at $T\leq T^*$ and $\rho=\rho_b=1-e^{-u/k_BT^*}$ at $T\geq T^*$.

In this model the phase transition depends on the presence of a box. 
The system is self-confined at any nonzero $T$ only if the box is large enough.
Raising $T$ expands the system.
The transition occurs, when further expansion is obstructed by the box.
Of greater interest is a case where the transition temperature depends only on the coupling strength of first-neighbor occupied cells.

\subsection{Logarithmic potential}\label{sec:log-pot}
The conditions for the occurrence of a phase transition at $T>0$ in $\mathcal{D}=1$ Ising models with long-range couplings have been studied thoroughly and produced an impressive collection of rigorous results, admirably summarized by Luijten and Bl\"ote \cite{LB97}.
For a power-law pair interaction potential, $J(r)\propto r^{-\alpha}$, the presence of a phase transition requires $\alpha\leq2$.
The coupling $J(r)$ in all these studies acts between all pairs Ising spins (or occupied cells).

In the context of our study, where the couplings are limited to first neighbors, no power-law potential, $\Phi(r)\propto r^{-\alpha}$ with $\alpha>0$, will produce a transition.
However, an attractive inverse-first-power interaction force between first-neighbor occupied cells will.
The associated logarithmic potential is transcribed into excitation energies for compacts of the form
\begin{equation}\label{eq:36} 
 \epsilon_m=mpV_c+u\ln (m+1)
\end{equation}

\begin{figure}[t]
  \begin{center}
 \includegraphics[width=75mm]{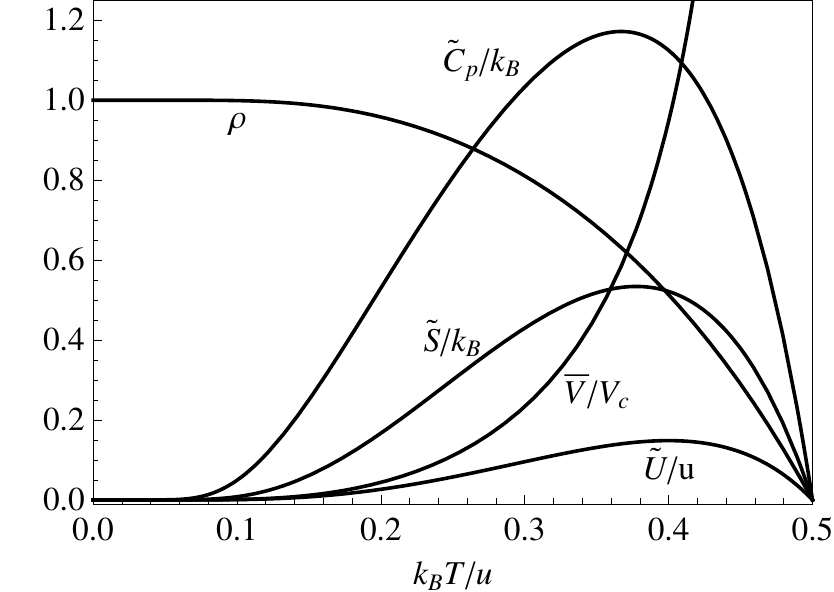}
\end{center}
\caption{Excess volume per occupied cell $\bar{V}/V_c$, fraction of occupied cells $\rho$, entropy $\tilde{S}/k_B$, heat capacity $\tilde{C}_p/k_B$, and internal energy $\tilde{U}/u$ (all per cell) as functions of scaled temperature $k_BT/u$ in the low-temperature phase at zero pressure.}
  \label{fig:fig4}
\end{figure}

The logarithmic interaction potential with $u>0$ supports a homogeneous phase at zero pressure and nonzero density of occupied cells as did the linear interaction potential considered in Sec.~\ref{sec:tri-wel}.
There, however, we found that when we increase $T$ the density $\rho$ decreases gradually and approaches zero asymptotically for $T\to\infty$.
Here we find that $\rho$ vanishes in a singularity at a finite and nonzero critical temperature $T_c$.
We can express the functions (\ref{eq:30}) for $p=0$ as follows:
\begin{align}\label{eq:38} 
& B_{00}(T,0)=\zeta(\beta u),\quad B_{10}(T,0)=\zeta(\beta u-1)-\zeta(\beta u), \nonumber \\
& B_{20}(T,0)=\zeta(\beta u-2)-2\zeta(\beta u-1)+\zeta(\beta u), \nonumber \\
& B_{01}(T,0)=-\beta u\,\zeta'(\beta u),\quad B_{02}(T,0)=(\beta u)^2\zeta''(\beta u), \nonumber \\
& B_{11}(T,0)=\beta u\big[\zeta'(\beta u)-\zeta'(\beta u-1)\big].
\end{align}
The mass density thus inferred from (\ref{eq:37}),
\begin{equation}\label{eq:52} 
 \rho=\frac{\zeta(\beta u)}{\zeta(\beta u-1)},
\end{equation}
approaches zero in a linear cusp singularity at temperature
\begin{equation}\label{eq:41} 
 k_BT_{c}=\frac{1}{2}u.
\end{equation}
The breakdown of self-confined extensivity at $T_c$ makes other thermodynamic quantities vanish in cusp singularities as well.
This is illustrated in Fig.~\ref{fig:fig4}.

The spacings distribution between occupied cells is qualitatively different for linear and logarithmic potentials.
It is exponential for the uniform attractive force: $\bar{N}_m\sim e^{-mu/k_BT}$.
The inverse first-power attractive force, by contrast, yields a power-law distribution: $\bar{N}_m\sim m^{-u/k_BT}$.
Raising the temperature broadens both distributions but in the latter case each moment diverges at some finite temperature.
The divergence of the first moment signals that $\rho$ approaches zero.
At $T>T_c$ the number of occupied cells increases sublinearly with volume on average.

\subsection{Continuum limit}\label{sec:con-lim}
Takahashi's exact solution for a gas of particles with generic first-neighbor interaction on a continuum  \cite{Taka42, BMD00} can be recovered by starting from expression (\ref{eq:22}) rewritten in the form
\begin{equation}\label{eq:40} 
 \beta G=-N_A\ln\left(1+\sum_{m=1}^\infty e^{-\beta[mpV_c+\phi_m-\phi_0]}\right).
\end{equation}
Setting $r=mV_c$, using $\phi_m=\Phi(mV_c)$ with $\phi_0=0$, and replacing the sum $\sum_m$ by the integral $V_c^{-1}\int dr$ we obtain
\begin{equation}\label{eq:46} 
 \beta G=-N_A\ln\left(\int_0^\infty \frac{dr}{V_c}\,e^{-\beta[pr+\Phi(r)]}\right).
\end{equation}
This differs from Takahashi's function $\Psi^*$ merely by the factor $V_c$ for length scale.

In the derivation of the equation of state and the entropy for the ideal Takahashi gas (ITG) from Eq.~(\ref{eq:46}) we must distinguish the volume $V_c$ of a vacant cell and the volume $V_A$ of an occupied cell, hitherto assumed equal.
The ITG expressions inferred from (\ref{eq:46}), 
\begin{equation}\label{eq:61} 
 \frac{pV_A}{k_BT}=\frac{\rho}{1-\rho},\quad 
 \frac{\tilde{S}}{k_B}=\rho\left[ 1+\ln\left(\frac{1-\rho}{\rho}\frac{V_A}{V_c}\right)\right],
\end{equation}
differ significantly from the corresponding ILG results presented in Sec.~\ref{sec:cont}.

\subsection{Hard rods with contact interaction}\label{sec:ha-ro-co-in}
Our vacancy particle approach is naturally generalizable to systems of hard rods with size $V_A=\sigma V_c$, $\sigma=1,2,\ldots$.
The first relation in Eq.~(\ref{eq:37a}) must then be replaced by
\begin{equation}\label{eq:29} 
 \sigma\left(\frac{1}{\rho}-1\right)= \frac{B_{10}(T,p)}{B_{00}(T,p)}.
\end{equation}

For rods with contact interaction $u$, the equation of state inferred from (\ref{eq:29}) then reads
\begin{equation}\label{eq:44} 
\rho=\frac{\sigma (x_p-1)[x_u(x_p-1)+1]}{\sigma (x_p-1)[x_u(x_p-1)+1]+x_p}.
\end{equation}
The entropy expression (\ref{eq:31}) holds for all $\sigma$.
These expressions coincide with our previous results derived by some of us via the completely different density functional approach \cite{BMM13}.

The vacancy particle approach covers new ground when we explore the impact of the contact interaction in the continuum limit.
The limit process is subtle.
Keeping the contact interaction strength $u$ fixed and taking the limit $V_c\to0$, $\sigma\to\infty$ at fixed $V_A=\sigma V_c$  leads to the ITG results (\ref{eq:61}).
Unsurprisingly, the lattice corrections fade away more rapidly for a repulsive contact interaction than for an attractive contact interaction as $V_c\to0$.

\begin{figure}[h]
  \begin{center}
 \includegraphics[width=41.5mm]{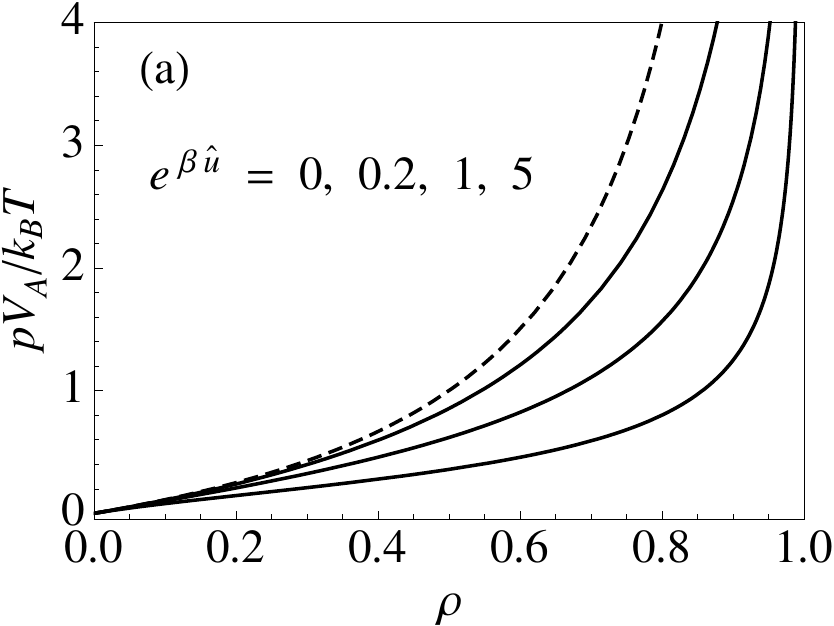}\hspace{2mm}%
 \includegraphics[width=42.5mm]{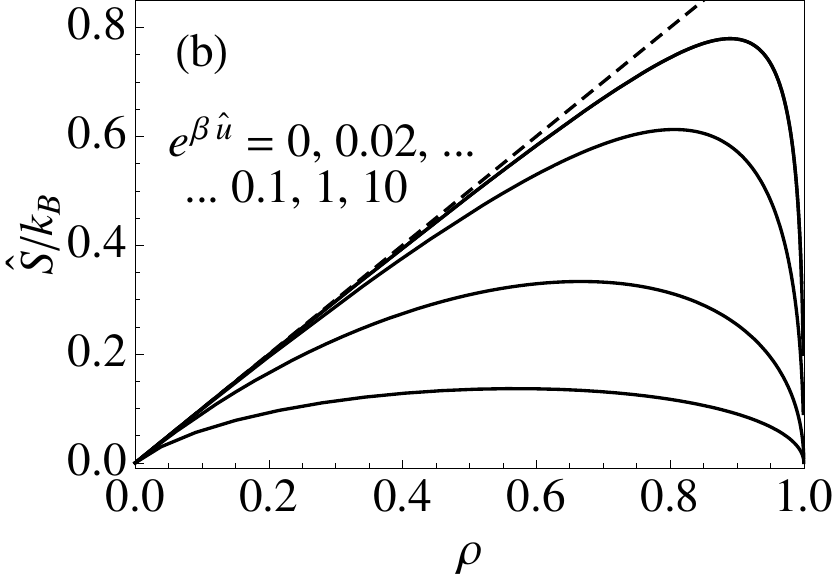}
\end{center}
\caption{(a) Equation of state and (b) entropy versus density for sticky hard rods in a continuum. The dashed lines represent the ITG gas.}
  \label{fig:fig8}
\end{figure}

A repulsive contact interaction of infinite strength $(u\to-\infty)$ will not survive the continuum limit either.
For an attractive contact interaction we consider the continuum limit $u\to\infty$, $\sigma\to\infty$ with fixed $e^{\beta\hat{u}}\doteq e^{\beta u}/\sigma$.
The ITG results emerge for $e^{\beta\hat{u}}=0$, whereas $e^{\beta\hat{u}}=\infty$ produces a single cluster as the only equilibrium state.
For finite and nonzero $e^{\beta\hat{u}}$ equilibrium states exist with the following parametric representations of the equation of state and the scaled entropy:
\begin{equation}\label{eq:62} 
\rho=\frac{\hat{K}_p^2e^{\beta\hat{u}}+\hat{K}_p}{\hat{K}_p^2e^{\beta\hat{u}}+\hat{K}_p+1},\quad 
\hat{K}_p\doteq\beta pV_A,
\end{equation}
\begin{equation}\label{eq:63} 
 \frac{\hat{S}}{k_B}=\lim_{\sigma\to\infty}\frac{\tilde{S}/k_B}{\ln\sigma}
 =\frac{\rho}{\hat{K}_pe^{\beta\hat{u}}+1}.
\end{equation}
Plots of $\hat{K}_p=pV_A/k_BT$ and $\hat{S}/k_B$ as functions of $\rho$ are shown in Fig.~\ref{fig:fig8} for various values of $e^{\beta\hat{u}}$.

%
\section{Conclusions and Outlook}\label{sec:ext}
%
The main goal of this work has been the introduction of a new methodology to the statistical mechanical analysis of lattice gas models in $\mathcal{D}=1$ dimensions with first-neighbor interactions of short and long range.
This includes derivative models such as hard rods of given size on a lattice or a continuum.

The methodology is an adaptation of fractional exclusion statistics as developed for quantum many-body systems \cite{Hald91a, Wu94, Isak94, Anghel}. 
In that quantum context Eq.~(\ref{eq:1}) plays the role of a generalized Pauli principle.
The same principle is applicable to the combinatorics of classical quasiparticles \cite{LVP+08, copic, picnnn, pichs, GKLM13}.

In the context of this work the statistically interacting quasiparticles are the vacancies between the material particles (atoms, rods etc.)
The interaction energies between the latter are encoded in the thermal excitation energies of the former.

The diverse applications are meant to establish the versatility of the approach.
Whereas all these applications have been limited to spatially homogeneous systems at thermal equilibrium and to interactions between first-neighbor material particles of arbitrary range and shape, such limitations are not intrinsic to the method.
Extensions to overcome one or the other limitation are in the works.

The inclusion of interaction potentials beyond first-neighbor material particles requires a significant extension of the analytic tools laid out in Secs.~\ref{sec:comb} and \ref{sec:stat-mech}.
Efforts in that direction have been carried out successfully for a different model \cite{picnnn}.
These extensions are adaptable to lattice gas models.

Contact has already been made (in Sec.~\ref{sec:ha-ro-co-in}) with another general method that has been employed with considerable success in the same arena: the density functional approach \cite{BMD00,BMM13}.
This method produces exact density functionals (of the material particles) for specific interactions and arbitrary external potentials.

The two methods have somewhat complementary strengths.
The vacancy particle approach produces results for longer-range potentials with greater ease. 
It also provides a rigorous framework for the calculation of the size distribution of vacant space between atoms.
The density-functional approach of Ref.~\cite{BMM13} is more readily amenable to the study of boundary effects and spatial correlations.
A comparative study of both approaches to interacting hard rods in a heterogeneous environment is in progress.

The study of polydisperse hard rods in homogeneous and heterogeneous media is another area for which both the vacancy particle and density functional approaches are well positioned as documented in prior work \cite{BSM12,LVP+08}
The development of the two methods in tandem to rods of mixed sizes opens up a host of applications that are likely to draw strong interest in the field of soft condensed matter physics.

The extension of the vacancy particle approach to nonequilibrium kinetics is a further challenge that presents itself naturally in the wake of this study.
There already exist advances in this direction that use density functional approaches  \cite{Dierl/etal:2012}.

\appendix

%
\section{Evaluation of partition function}\label{sec:parti}
%
The derivation of Eqs.~(\ref{eq:12})-(\ref{eq:14}) from expression (\ref{eq:11}) with ingredients (\ref{eq:2}) and (\ref{eq:3}) was given by Wu \cite{Wu94}.
It is summarized here for completeness.
We avoid scaling by capacity constants $A_m$.
Vanishing $A_m$ do exist as, for example, in (\ref{eq:5}).

In a macroscopic system, the partition function (\ref{eq:11}) is very sharply peaked for the most probable populations $\hat{N}_m$ of particles from all species $m$, dominated by a single term,
\begin{equation}\label{eq:a1} 
Z=W(\{\hat{N}_m\})\exp\left(-\beta E(\{\hat{N}_m\})\right)\Big[1+\ldots\Big].
\end{equation}
By using Stirling asymptotics, $\ln(\hat{N}_m!)=\hat{N}_m\ln\hat{N}_m-\hat{N}_m+\mathrm{O}(\ln \hat{N}_m)$, and dropping terms of O$(\ln \hat{N}_m)$ one obtains
\begin{align}\label{eq:a2} 
\ln Z &= \sum_m\Big[\Big(A_m+\sum_{m'}(\delta_{mm'}-g_{mm'})
\hat{N}_{m'}\Big)\nonumber \\
&\hspace*{15mm}\times\ln\Big(A_m+\sum_{m'}(\delta_{mm'}-g_{mm'})
\hat{N}_{m'}\Big) \nonumber \\
&-\Big(A_m-\sum_{m'}g_{mm'}
\hat{N}_{m'}\Big)\ln\Big(A_m-\sum_{m'}g_{mm'}\hat{N}_{m'}\Big) \nonumber \\
&\hspace*{25mm}-\hat{N}_m\ln\hat{N}_m -\beta\hat{N}_m\epsilon_m\Big].
\end{align}
The extremum condition, $\partial\ln Z/\partial\hat{N}_m=0$, leads to the following coupled equations for the $\hat{N}_m$:
\begin{align}\label{eq:a3} 
\ln\hat{N}_m+\beta\epsilon_m &= \ln\Big(A_m+\sum_{m''}(\delta_{mm''}-g_{mm''})
\hat{N}_{m''}\Big) \nonumber \\
&\hspace{-5mm}+\sum_{m'}g_{m'm}\Big[\ln\Big(A_{m'}-\sum_{m''}g_{m'm''}\hat{N}_{m''}\Big) 
\nonumber \\ &\hspace{-10mm}-\ln\Big(A_{m'}+\sum_{m''}(\delta_{m'm''}-g_{m'm''})\hat{N}_{m''}\Big)\Big]
\end{align}
The solution of Eqs.~(\ref{eq:a3}) is worked out in two steps by
introducing the following quantities:
\begin{equation}\label{eq:a4} 
w_m\doteq \frac{A_m}{\hat{N}_m}-\sum_{m'}g_{mm'}\frac{\hat{N}_{m'}}{\hat{N}_m}.
\end{equation}
Substitution of (\ref{eq:a4}) into an exponentiated version of (\ref{eq:a3}) yields coupled equations for the $w_m$ that have a simpler structure, namely Eqs.~(\ref{eq:13}). 
The $\hat{N}_m$ are then recovered from Eqs.~(\ref{eq:a4}), which transcribe into Eqs.~(\ref{eq:14}) for the population densities $\bar{N}_m$.
The partition function (\ref{eq:a2}) expressed as a function of the $w_m$ acquires the simple form (\ref{eq:12}).
The entropy expression (\ref{eq:17}) follows from $S=k_B\ln W(\{\hat{N}_m\})$ using the right-hand side of (\ref{eq:a2}) without the last term.
It can be rendered more compactly in the form,
\begin{equation}\label{eq:a5} 
\frac{\bar{S}}{k_B}=\sum_m\bar{N}_m\big[(1+w_m)\ln(1+w_m)-w_m\ln w_m\big],
\end{equation}
where the $w_m$ from (\ref{eq:a4}) can be interpreted, in conjunction with (\ref{eq:3b}), as the number of open slots to place particles from species $m$, scaled by the number of such particles already in the macrostate.
When a population $\hat{N}_m$ is depleted or frozen out by a particular choice of control variables or parameters, the associated $w_m$ diverges, thus removing its presence in the partition function (\ref{eq:12}).

In applications where population densities $\bar{N}_m$ are controlled by chemical potentials, all occurrences of $\epsilon_m$ must be replaced by $\epsilon_m-\mu_m$.

%
\section{Correlations of occupied cells}\label{sec:appb}
%
Correlations of first-neighbor occupied cells belong to the primary target quantities in the density functional approach \cite{BMM13}, which is capable of producing exact results for much the same interaction potentials.
Here we explain the relations between those correlations and the distribution of (compact) vacancy particles calculated in this work, with a simple case worked out for a guide.

Using the occupation number $\tau_i=0,1$ to indicate vacancy or occupancy and assuming homogeneity we write for the average occupancy,
\begin{equation}\label{eq:B1} 
 \rho\doteq\langle\tau_i\rangle=\frac{N_A}{N},
\end{equation}
where $N_A$ is the number of occupied cells and $N$ the total number of cells, vacant or occupied.
Using the short-hand notation,
\begin{equation}\label{eq:B2} 
C_l\doteq\langle\tau_i\tau_{i+l}\rangle,\quad
C_{lk}\doteq\langle\tau_i\tau_{i+l}\tau_{i+l+k}\rangle,\quad \cdots
\end{equation}
for the $n$-point correlators of occupied cells (with $n=2,3,\ldots$), we can express the rescaled population densities of compact vacancy particles, $\tilde{N}_m\doteq\rho\bar{N}_m$, $m=1,2,\ldots$, as follows:
\begin{align}\label{eq:B3}
\tilde{N}_0 &=\langle\tau_1\tau_2\rangle=C_1, \\
\tilde{N}_1 &=\langle\tau_1(1-\tau_2)\tau_3\rangle=C_2-C_{11}, \nonumber 
\end{align}
\begin{align}
\tilde{N}_2 &=\langle\tau_1(1-\tau_2)(1-\tau_3)\tau_4\rangle=C_3-2C_{12}+C_{111}, \nonumber \\
\tilde{N}_3 &=\langle\tau_1(1-\tau_2)(1-\tau_3)(1-\tau_4)\tau_5\rangle \nonumber \\ &=C_4-2C_{22}-2C_{13}+2C_{112}+C_{121}-C_{1111}, \nonumber
\end{align}
etc.
The usefulness of these relations hinges on whether or not they can be inverted to extract the pair correlations $C_l$, $l=1,2,\ldots$ for occupied cells at a distance $lV_c$ from the population distribution $\tilde{N}_m$, $m=0,1,2,\ldots$ of compact vacancy particles with size $mV_c$.

In the following we consider the two simplest cases and show how to extract pair correlations,
\begin{equation}\label{eq:B4} 
 \hat{C}_l\doteq\langle\tau_i\tau_{i+l}\rangle-\langle\tau_i\rangle\langle\tau_{i+l}\rangle
 =C_l-\rho^2,
\end{equation}
for arbitrary distances from the relations (\ref{eq:B3}).

For the ILG (Sec.~\ref{sec:cont}) we have $\tilde{N}_m=\rho^2(1-\rho)^m$, where $\rho=1-x_p^{-1}$.
We infer $C_l=\rho^2$, $C_{lk}=\rho^3$, etc., implying identically vanishing correlations, $\hat{C}_l\equiv0$, at all distances $l$ as expected.

In the case of contact interactions we can reduce all $n$-point correlators to two-point correlators:
\begin{equation}\label{eq:B5} 
 C_{11}=\frac{C_1^2}{\rho},\quad C_{12}=\frac{C_1C_2}{\rho},\quad 
 C_{111}=\frac{C_1^3}{\rho^2},\quad \cdots,
\end{equation}
which then makes it possible to express any $C_l$ as a function of the $\tilde{N}_m$ for $m=0,1,\ldots,l-1$:
\begin{align}\label{eq:B6}
C_1 &=\tilde{N}_0,\qquad C_2 =\tilde{N}_1+\frac{\tilde{N}_0^2}{\rho}, \nonumber \\
C_3 &=\tilde{N}_2+2\frac{\tilde{N}_0\tilde{N}_1}{\rho}+\frac{\tilde{N}_0^3}{\rho^2},\quad\ldots
\end{align}

%
%


\end{document}